\documentstyle[aps]{revtex}
\begin{document}

\draft
\title{
Friedel Oscillations in the One-Dimensional Kondo Lattice Model
}
\author{Naokazu Shibata} 
\address{
Institute for Solid State Physics, University of Tokyo,  
7-22-1 Roppongi, Minato-ku, Tokyo 106, Japan, \\
and Department of Physics, Faculty of Science,
Science University of Tokyo, 
1-3 Kagurazaka, Shinjuku-ku, Tokyo 162, Japan 
}
\author{Kazuo Ueda}
\address{
Institute for Solid State Physics, University of Tokyo, 
7-22-1 Roppongi, Minato-ku, Tokyo 106, Japan
}
\author{Tomotoshi Nishino}
\address{
Department of Physics, Graduate School of Science, 
Tohoku University, Sendai 980, Japan
}
\author{Chikara Ishii}
\address{
Department of Physics, Faculty of Science, 
Science University of Tokyo, 
1-3 Kagurazaka, Shinjuku-ku, Tokyo 162, Japan
}
\date{\today}
\twocolumn

\maketitle
\begin{abstract}
The paramagnetic metallic  phase of the one-dimensional Kondo 
lattice model is studied by the density matrix 
renormalization group method. We observe charge and spin 
Friedel oscillations. They reflect the long range charge-charge
and spin-spin correlation functions. The observed oscillations 
are consistent with a Tomonaga-Luttinger liquid. 
From the period of the oscillations it is concluded that the 
Fermi surface is large, including both the conduction electrons 
and the localized spins, $k_F=\pi (1+n_c)/2$, where $n_c$ is the 
density of conduction electrons.
\end{abstract}

Electron density oscillations as a response to a local
perturbation provide a clue to understanding electronic states 
of a system. Response to an impurity potential is known
as Friedel oscillations. The period of the 
oscillations is given by the diameter of the Fermi surface which 
is defined by the singular points in the momentum distribution
function. As a local perturbation it is also possible to use a 
magnetic impurity. Then spin density oscillations are induced.
Since these oscillations are the origin of the 
Ruderman-Kittel-Kasuya-Yosida (RKKY) interaction,
the Friedel oscillations are responsible for the magnetically 
ordered structure of rare earth metals.

Generally speaking, a key to understanding the physics of
heavy fermion systems is the competition between the 
RKKY interaction and the Kondo screening effect of the 
magnetic moments. 
The Kondo lattice model is one of the canonical models for
heavy fermions and much effort has been devoted to elucidate 
the properties of the model.
Recently the ground state phase diagram of the Kondo lattice model
was completed in one-dimension\cite{1D,ED}. 
There are three phases in one dimension: 
a ferromagnetic metallic, a paramagnetic
metallic, and an insulating spin liquid phase.

Our concern in the present paper is the metallic phase of the
one-dimensional Kondo lattice model. In one dimension it has been 
established that most interacting metallic systems belong to the
universality class of Tomonaga-Luttinger liquids\cite{haldane}.
The asymptotic forms of charge- and spin-correlation functions are 
\begin{eqnarray}
\langle n(x)n(0)\rangle&=&K_\rho/(\pi x)^2+A_1\cos(2k_Fx)
x^{-1-K_\rho}  \nonumber \\ 
 & & + A_2\cos(4k_Fx)x^{-4K_\rho} , \\
\langle S(x)\cdot S(0)\rangle&=& 
1/(\pi x)^2+B_1\cos(2k_Fx)x^{-1-K_\rho},
\end{eqnarray}
where $k_F=\pi\rho/2$, with $\rho$ being the density of charge 
carriers, is the Fermi momentum and
$K_\rho$ is the correlation exponent\cite{schulz}. 
In the above equations logarithmic 
corrections to the $2k_F$ correlations 
have been omitted. The momentum distribution function around $k_F$ 
shows a power law singularity $n_k \sim 1/2 - sgn (k-k_F)
|k-k_F|^\alpha$ with $\alpha=(K_\rho+1/K_\rho-2)/4$.
The anomalous power law decays of the correlation functions naturally
reflect themselves in the Friedel oscillations:
the asymptotic form of the charge density oscillations 
induced by an impurity potential is
\begin{eqnarray}
\delta \rho(x)& \sim & C_1\cos(2k_Fx)x^{(-1-K_\rho)/2}
 + C_2\cos(4k_Fx)x^{-2K_\rho}
\end{eqnarray}
as a function of the distance $x$ from the impurity
\cite{wire1,wire2,wire3}.
Analogously, the spin density oscillations 
induced by a local magnetic field behave as 
\begin{eqnarray}
 \sigma(x) & \sim & D_1\cos(2k_Fx)x^{-K_\rho}.
\end{eqnarray}

Concerning the one-dimensional Kondo lattice model the paramagnetic
metallic state is expected to belong to the class of Luttinger 
liquids. However in this case the position of the Fermi points 
is already a nontrivial problem since the model
consists of two components with completely different characters, 
the conduction electrons and the localized spins.

There are two different points of view concerning the above question.
If the interaction between the conduction electrons 
and the localized spins is strictly zero, it is clear that the 
singularity in the momentum distribution
function of the conduction electrons is determined only by 
the number of conduction electrons.
Thus, if the singular points are not affected by  
the interaction, $k_F=\pi n_c /2$ is expected.
On the other hand, a different answer is obtained 
by identifying the Kondo lattice model as an effective model
for the periodic Anderson model. 
In the periodic Anderson model the conduction electrons 
and the $f$ electrons are mixed with each other through
the hybridization matrix elements.
Therefore it is naturally expected that $k_F$ 
is determined by the total density of both the conduction 
electrons and the $f$ electrons; $k_F=\pi(1+n_c)/2$.

In order to draw a conclusion on whether the Luttinger sum rule 
includes localized spins or not, it is necessary to deal with 
large-size systems because the Luttinger liquid is characterized 
by the long range correlations and the singularity in the 
momentum distribution function is clearly defined 
only in the infinite system. For this purpose the density matrix 
renormalization group (DMRG) method developed by White\cite{DMRG} 
is most promising since in the DMRG we can study long chains and 
obtain results with only small systematic errors, 
which can be estimated from the eigenvalues of 
the density matrix.

In this paper we calculate the Friedel oscillations of the
one-dimensional Kondo lattice model by using the DMRG. 
The charge density oscillations 
are induced naturally by open boundary conditions and 
the spin density oscillations are introduced  
by applying local magnetic fields at both ends.
From the period of the Friedel oscillations
we can determine the Fermi momentum.
Surprisingly the Friedel oscillations are compatible with the large
Fermi surface $k_F=\pi(1+n_c)/2$
in spite of the fact that the charge degrees of freedom
are completely suppressed for the $f$ electrons in the Kondo lattice
model. 

The Hamiltonian we use in the present study is the usual 
one-dimensional Kondo lattice model,
\begin{equation}
H  =  -t\sum_{i \sigma}
( c_{i \sigma}^\dagger c_{i+1 \sigma} + \mbox{H.c.})  
+J \sum_{i \mu} S^\mu_i s^\mu_i 
\end{equation}
where $c_{ i \sigma}^\dagger \  (c_{ i \sigma}) $ is the creation 
(annihilation) operator of a conduction electron  
at the $i$-th site, and $ s^\mu_i=(1/2)\sum_{\sigma\sigma'}
c_{i\sigma}^{\dagger} \tau^\mu_{\sigma\sigma'}c_{i\sigma'} $,
with $\tau^\mu_{\sigma\sigma'}$ ($\mu = x,y,z$) being the 
Pauli matrices, are the spin density operators of the conduction 
electrons. The spin densities
are coupled to the localized spins $S^\mu_i$ through an 
antiferromagnetic exchange coupling $J$.

Here we briefly summarize the phase diagram 
of this model\cite{1D,ED}. 
At half filling, $n_c=1$, the ground state is a spin singlet
with a gap for excitations\cite{gap,yu,tvlk,KLU}. 
Away from half filling the ground state is ferromagnetic
in the strong coupling limit\cite{troy,ferr}. 
The ferromagnetic ground state is continuously connected to the
low carrier density limit, $n_c \rightarrow 0$,
where the ferromagnetic state survives even in the limit of 
weak exchange couplings.
The paramagnetic metallic ground state exists only for weak 
couplings for finite carrier densities.
The phase boundary between the ferromagnetic state and the
paramagnetic state had been determined by the numerical exact
diagonalization of finite clusters, and it was shown 
that the critical value $J_c$ goes up with increasing
the carrier density\cite{ED}.  

To treat large systems with sufficient 
accuracy, small truncation errors in the DMRG calculation are 
necessary.
Since the truncation errors decrease with increasing $J$,
we chose $n_c=4/5$ for the calculation of Friedel oscillations,
where the paramagnetic state remains up to $J_c = 3.0t$.
At this filling there is a small additional
ferromagnetic region below $J_c=3.0t$. It has already 
been reported in previous exact diagonalization 
studies at slightly different carrier concentration
$n_c=0.75$\cite{ED}. Our DMRG results for clusters of 
size $N=10$ and $20$ confirm the existence of
this ferromagnetic region 
between $J = 1.6t$ and $J = 1.8t$ at $n_c=4/5$.
The total spin of the ferromagnetic state is the same as 
that for strong coupling, and 
the phase transitions are caused by simple level 
crossings between the two lowest states with $S=0$ and
$S=N(1-n_c)/2$.
We calculate Friedel oscillations in the two paramagnetic 
states at $J=2.5t$ and $J=1.5t$.

We first consider the paramagnetic state at $J=2.5t$.
The charge density oscillations induced by the open boundary 
conditions are presented in Fig. 1 by the solid line. 
As expected, long range oscillations characteristic of 
a Luttinger liquid are induced by the boundary conditions.
Since the weak decay of the oscillations makes it difficult to 
determine the correlation exponents $K_\rho$ by the present 
system size, we focus our attention mainly on the period of the 
oscillations. 
The Fourier components of the oscillations, presented in 
Fig. 2, show a clear single peak at $q=2\pi/5$. 
The single peak at this wave number is natural in 
the strong coupling limit where the conduction electrons and 
the localized spins form local singlets, leading to complete 
spin-charge separation.
In this case we can treat the charge part as spinless fermions
whose correlation functions are characterized by its single 
$2k_F$ structure given by $2\pi(1-n_c)$. 
The peak at $q=2\pi/5$ corresponds to the 
$2k_F$ structure of the spinless fermions which is equivalent to 
$4k_F$ structure of the original fermions.
The result of Fig. 1 means that the charge density oscillations 
at $J=2.5t$ are already well characterized by the nature of the 
strong coupling region.

From the above results,
it is not possible to draw a conclusion about whether the 
Fermi surface includes the localized spins, because 
$4k_F=2\pi (1+n_c) = 2\pi n_c\ (mod\ 2\pi)$.  
In order to see the contribution of the localized spins
we next calculate spin density oscillations.
If the $2k_F$ structure is found in the spin density 
Friedel oscillations reflecting the spin-spin correlations 
of Luttinger liquids, it may be possible to argue 
whether the Fermi surface includes the localized spins or not: 
if the localized spins contribute to the Fermi surface a 
$2k_F=\pi/5$ structure should appear but if it does not
a $4\pi/5$ structure appears. 
To induce the Friedel oscillations of spin density, 
we apply local magnetic fields, 
$H_{local}=2h(S^z_1 - s^z_1 -S^z_N + s^z_N)$
to the spins and conduction electrons at the boundary sites. 
These local magnetic fields with opposite directions for the
two boundary spins induce oscillations that are odd with 
respect to reflection. This feature has been used to confirm
the convergence of the DMRG.

The spin density oscillations are presented
in Fig.\ 3 by the solid line and their Fourier components are
shown in Fig.\ 4. 
We clearly see long range oscillations with $q=\pi/5$.
They are consistent with a Luttinger liquid prediction, 
Eq.\ (4), if we assume the large Fermi surface where
both the conduction electrons and the localized spins
contribute to the Fermi volume:
\begin{equation}
k_F  = \frac{\pi}{2} ( 1 + n_c ) .
\end{equation}

Now we proceed to the paramagnetic metallic phase in the
weak coupling region.
The result of the charge density oscillations 
at $J=1.5t$ is presented in Fig.\ 1 by the broken line.
In this case too long range oscillations are induced
and their period is the same as $J=2.5t$.
The similarity is also seen in the Fourier components 
shown in Fig.\ 2 by the broken line.  
For these coupling strengths we cannot see a peak at 
$q=2k_F=\pi/5$.
This means the amplitude of the $2k_F$ oscillations of the
charge density is still negligible even in the case of $J=1.5t$. 
However, as is shown in the inset of Fig.\ 2 
we find clear $2k_F$ oscillations in addition to the 
dominant $4k_F$ oscillations
at a smaller coupling $J=1.0t$ which is consistent with the 
general form of the Luttinger liquids.

For the spin density at $J=1.5t$, the oscillations and their 
Fourier components are presented in Fig.\ 3 and Fig.\ 4, 
respectively, by the broken lines.
Although a small structure is also found at $q=3\pi/5$, 
the dominant component is at $q=\pi/5$.
The structure at $q=3\pi/5$ is considered to be induced by
the coupling mode of the spin density oscillations with 
$q=\pi/5$ and the charge density oscillations with $q=2\pi/5$ 
whose amplitude is larger than that of $J=2.5t$.
Since the structure at $q=3\pi/5$ decreases rapidly 
with an increasing system size compared with 
the structure at $q=\pi/5$, we observe that it is not 
an intrinsic property of the infinite system. 
Combined with the results of the charge density oscillations
it is concluded that the paramagnetic state at $1.5t$ is 
also characterized by the Luttinger liquid with the large 
Fermi surface.

At representative values for the exchange coupling constant 
$J=2.5t$ and $J=1.5t$, we have seen that the paramagnetic metallic 
phases of the Kondo lattice model show Friedel oscillations
characteristic of a Luttinger liquid with a large Fermi
surface. Similar results are obtained at different concentrations
of conduction electrons, $n_c=2/3,J=2.0t$ and $n_c=6/7, J=1.7t$.
These results are consistent with the previous 
work on the $t$-$t'$ Kondo lattice model for which it was 
shown exactly that its strong coupling limit is described 
by a Luttinger liquid with the large Fermi surface\cite{tt'}. 
Therefore it is natural to
conclude that the paramagnetic metallic phase of the
Kondo lattice model has a large Fermi surface in 
general. This conclusion is consistent with a variational
Monte Carlo study using Gutzwiller projected 
hybridization form\cite{VMC}.
This conclusion is also consistent with the bosonization study
by Fujimoto and Kawakami\cite{fujimoto}. 
Although their work was recently criticized by White and 
Affleck\cite{zigzag}, it should be noted that the model
used by the latter authors is different from the usual
Kondo lattice model.

The DMRG method has made it possible for the first time to
observe the Friedel oscillations corresponding to the large
Fermi surface in the paramagnetic phase of the weak coupling 
region. However, it is still difficult to
see them in the weak coupling limit $J<t$, where the
length scale of the Luttinger liquid properties
clearly exceeds the system size studied in the present
work. Calculations on longer systems and the determination of 
the correlation exponent $K_\rho$ are important areas for future 
investigation.

%\acknowledgments
A part of the numerical calculation was done by VPP500
at the Supercomputer Center of the ISSP, University of Tokyo.
We thank Matthias Troyer for helpful discussions and
a critical reading of the manuscript.
This work is financially supported by Grant-in-Aid from 
the Ministry of Education, Science, Sports and Culture of Japan.
N.\ S.\ is supported by the Japan Society for the 
Promotion of Science.

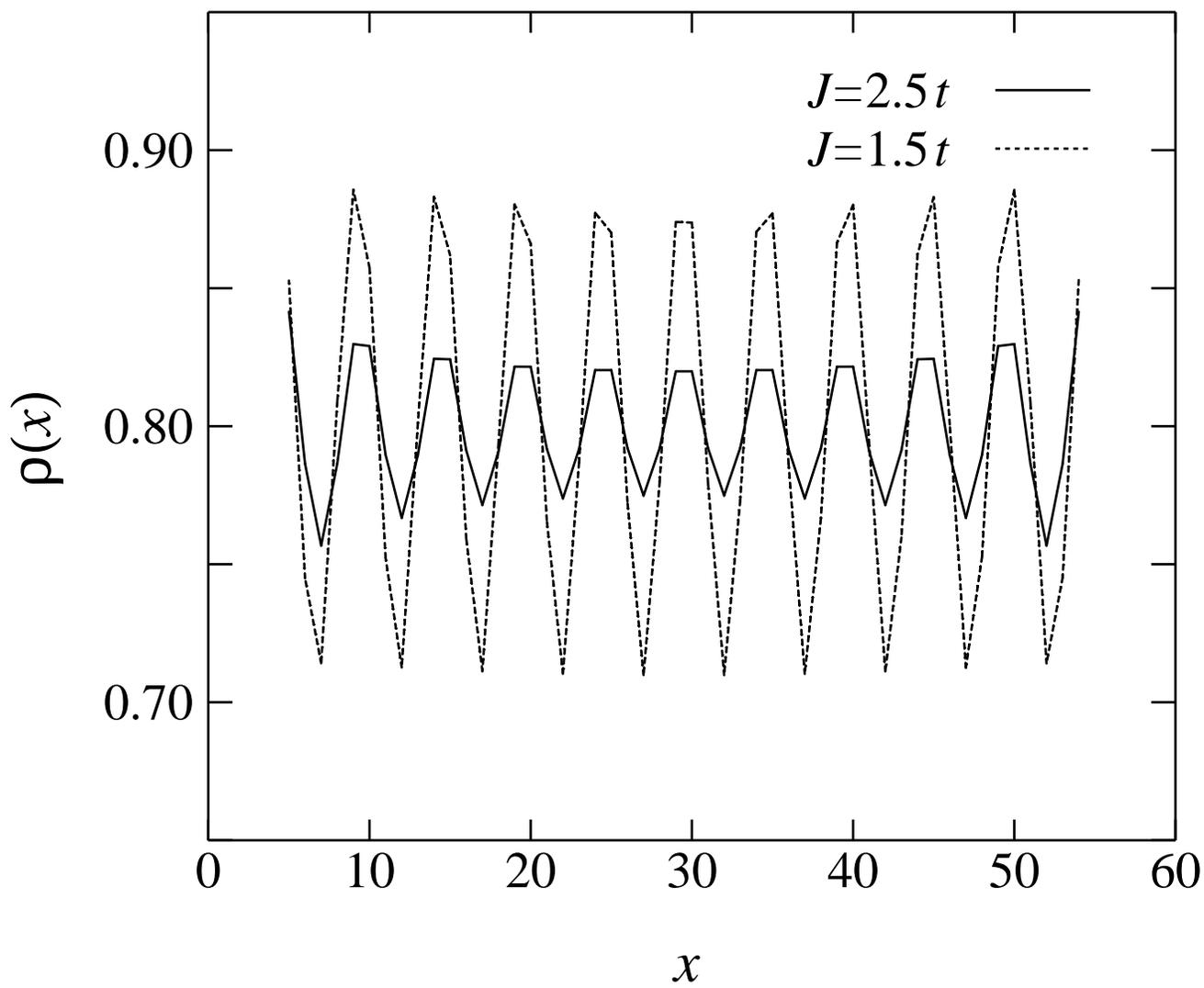
\begin{figure}
\caption{
Charge density oscillations of the Kondo lattice model.
The system size is 60 sites and the carrier density is $n_c=4/5$.
The solid line and the broken line correspond to $J=2.5t$
and $J=1.5t$, respectively. 
Typical truncation errors in the DMRG calculations 
are $1 \times 10^{-6}$ for $J=2.5t$
and $3 \times 10^{-6}$ for $J=1.5t$.
}
\label{charge-r}
\end{figure}

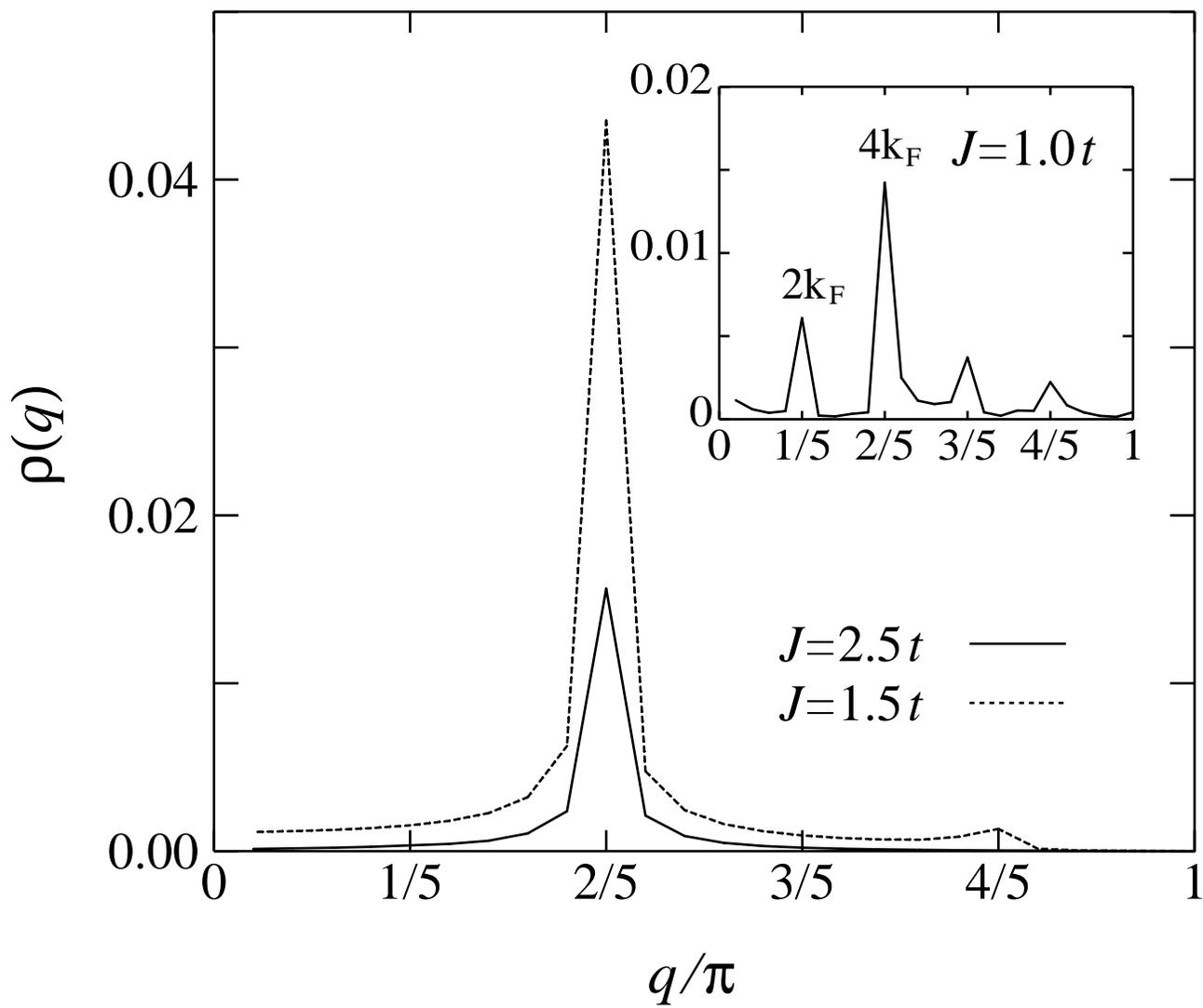
\begin{figure}
\caption{
Fourier components of the charge density oscillations.
The system size is 60 sites and the carrier density is $n_c=4/5$.
The solid line and the broken line 
correspond to $J=2.5t$ and $J=1.5t$, respectively.
The Fourier transformation is carried out by using 
the central 50 sites. The inset shows the result for $J=1.0t$, 
for which typical truncation errors are $4 \times 10^{-5}$.
}
\label{charge-q}
\end{figure}

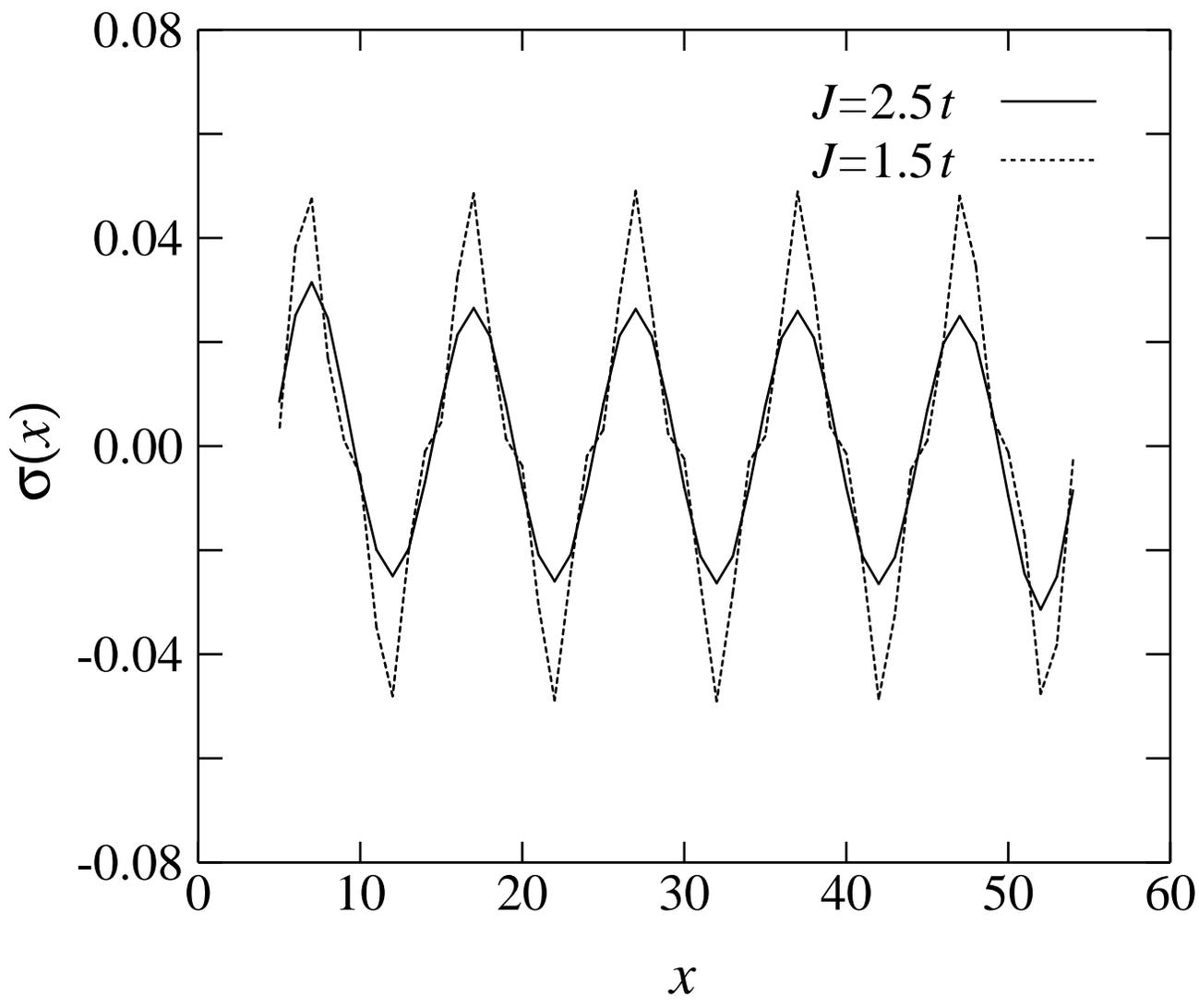
\begin{figure}
\caption{
Spin density oscillations of the Kondo lattice model.
The system size is 60 sites and the carrier density is $n_c=4/5$.
The solid line and the broken line correspond to $J=2.5t$
and $J=1.5t$, respectively. The strength of the local magnetic 
field $h$ is $0.1t$.
Typical truncation errors in the DMRG calculations 
are $1 \times 10^{-6}$ for $J=2.5t$ 
and $3 \times 10^{-6}$ for $J=1.5t$.
}
\label{spin-r}
\end{figure}
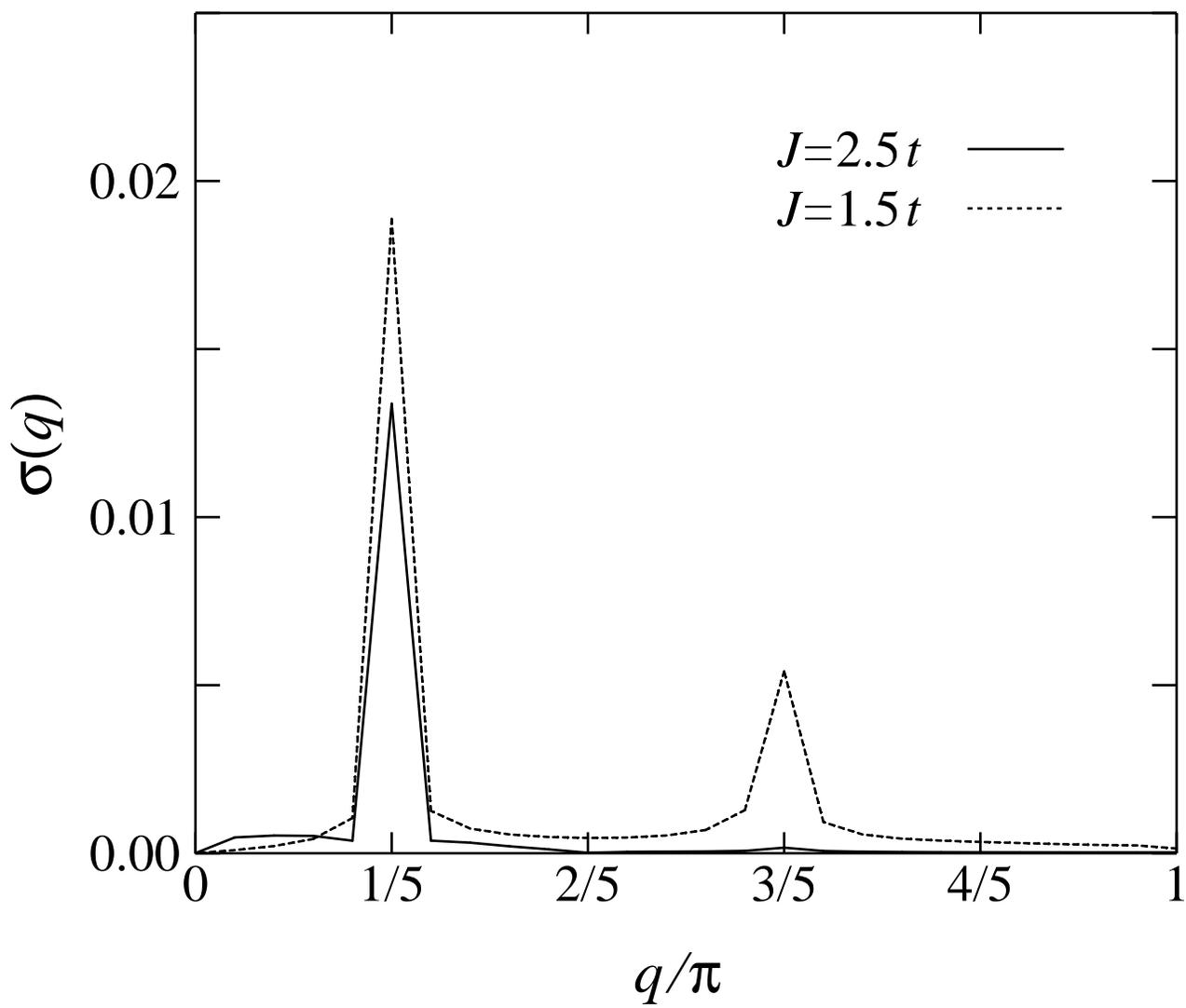
\begin{figure}
\caption{
Fourier components of the spin density oscillations.
The system size is 60 sites and the carrier density is $n_c=4/5$.
The solid line and the broken line correspond to $J=2.5t$
and $J=1.5t$, respectively.
The Fourier transformation is carried out by using 
the central 50 sites.
}
\label{spin-q}
\end{figure}

\end{document}